\documentclass[iop,apj,numberedappendix]{emulateapj}
\usepackage[latin1]{inputenc}
\usepackage{amssymb}
\usepackage{amsmath}
\usepackage{graphicx}
\usepackage{xspace}
\usepackage{natbib}
\usepackage{url}
\usepackage{paralist}

\slugcomment{Accepted for publication in ApJ Letters}

\shorttitle{X-ray dips in Fairall~9}
\shortauthors{Lohfink et al.}

\begin{document}

\title{X-ray dips in the Seyfert Galaxy Fairall~9:\\ Compton-thick ``comets" or a failed radio galaxy?}


\author{Anne M. Lohfink\altaffilmark{1,2}, Christopher S. Reynolds\altaffilmark{1,2}, Richard F. Mushotzky\altaffilmark{1,2}, J\"orn Wilms\altaffilmark{3}}
\altaffiltext{1}{Department of Astronomy, University of Maryland, College Park, MD 20742-2421, USA; alohfink@astro.umd.edu}
\altaffiltext{2}{Joint Space-Science Institute (JSI), College Park, MD 20742-2421, USA}
\altaffiltext{3}{Dr. Karl-Remeis Observatory and ECAP, Sternwartstr. 7, 96049 Bamberg, Germany}


\begin{abstract}
\noindent We investigate the spectral variability of the Seyfert galaxy Fairall~9 using almost 6\,years of monitoring with the {\it Rossi X-ray Timing Explorer (RXTE)} with an approximate time resolution of 4 days.   We discover the existence of pronounced and sharp dips in the X-ray flux, with a rapid decline of the 2--20 keV flux of a factor 2 or more followed by a recovery to pre-dip fluxes after $\sim 10\,{\rm days}$.   These dips skew the flux distribution away from the commonly observed log-normal distribution. Dips may result from the eclipse of the central X-ray source by broad line region (BLR) clouds, as has recently been found in NGC~1365 and Mrk~766.  Unlike these other examples, however, the clouds in Fairall~9 would need to be Compton-thick, and the non-dip state is remarkably free of {\it any} absorption features.   A particularly intriguing alternative is that the accretion disk is undergoing the same cycle of disruption/ejection as seen in the accretion disks of broad line radio galaxies (BLRGs) such as 3C120 but, for some reason, fails to create a relativistic jet.  This suggests that a detailed comparison of Fairall~9 and 3C120 with future high-quality data may hold the key to understanding the formation of relativistic jets in AGN.
\end{abstract}



\section{AGN Variability}\label{intro}

X-rays are emitted from the inner accretion disks around both stellar-mass and supermassive black holes.   This gives us our most effective probe to date of accretion physics close to black holes as well as the immediate environments of black holes.   Studies of X-ray spectral variability are a particularly powerful diagnostic.  For example, combined spectral and timing analyses have led to the discovery of distinct hardness/flux states in X-ray binaries \citep{tanaka:96a} which almost certainly correspond to distinct accretion modes.  Spectral variability also facilitates the decomposition of a complex spectrum into physical components  that would be degenerate in a single spectrum.  

Detailed studies of long term (year$+$) X-ray variability in active galactic nuclei (AGN) have been possible since the launch of the {\it Rossi X-ray Timing Explorer} ({\it RXTE}) but a relatively small number of sources have been targeted for detailed study \citep{markowitz:03a,rivers:11a}.  These studies are crucially important as they probe timescales comparable to the viscous timescale on which the true mass accretion rate would be expected to vary.  More typical X-ray observations of active galactic nuclei (AGN) probe timescales of hours-to-days, characteristic of dynamical or thermal timescales but much shorter than the viscous timescales.   Long-term studies are also important for uncovering rare/transient phenomena that may give unique windows into the physics and structure of AGN.   

In this {\it Letter}, we report the discovery of an unusual pattern of X-ray variability in the nearby Seyfert-1 galaxy Fairall~9 ($z=0.047$) using the long-term monitoring data from pointed {\it RXTE} observations. Rather than displaying the usual ``flary" lightcurve with a log-normal flux distribution \citep{gaskell:03a,uttley:05a}, Fairall~9 shows long-timescale modulations punctuated by strong and sudden {\it dips}.  This behavior was apparent but not commented upon when the monitoring was first presented by \citet{markowitz:03a}.   Using the latest background models and calibration files, we analyze all available \textit{RXTE} data and confirm the presence of these dips in the \textit{RXTE} lightcurve.  We discuss their possible origins, including the possibility that they correspond to eclipses of the X-ray source by clouds in the broad line region (BLR).   Alternatively, the dips may result from a genuine accretion disk instability, suggesting an interesting parallel between Fairall~9 and broad line radio galaxies.  

\section{Observations}

\begin{figure*}[t]
\hbox{
\includegraphics[width=0.5\textwidth]{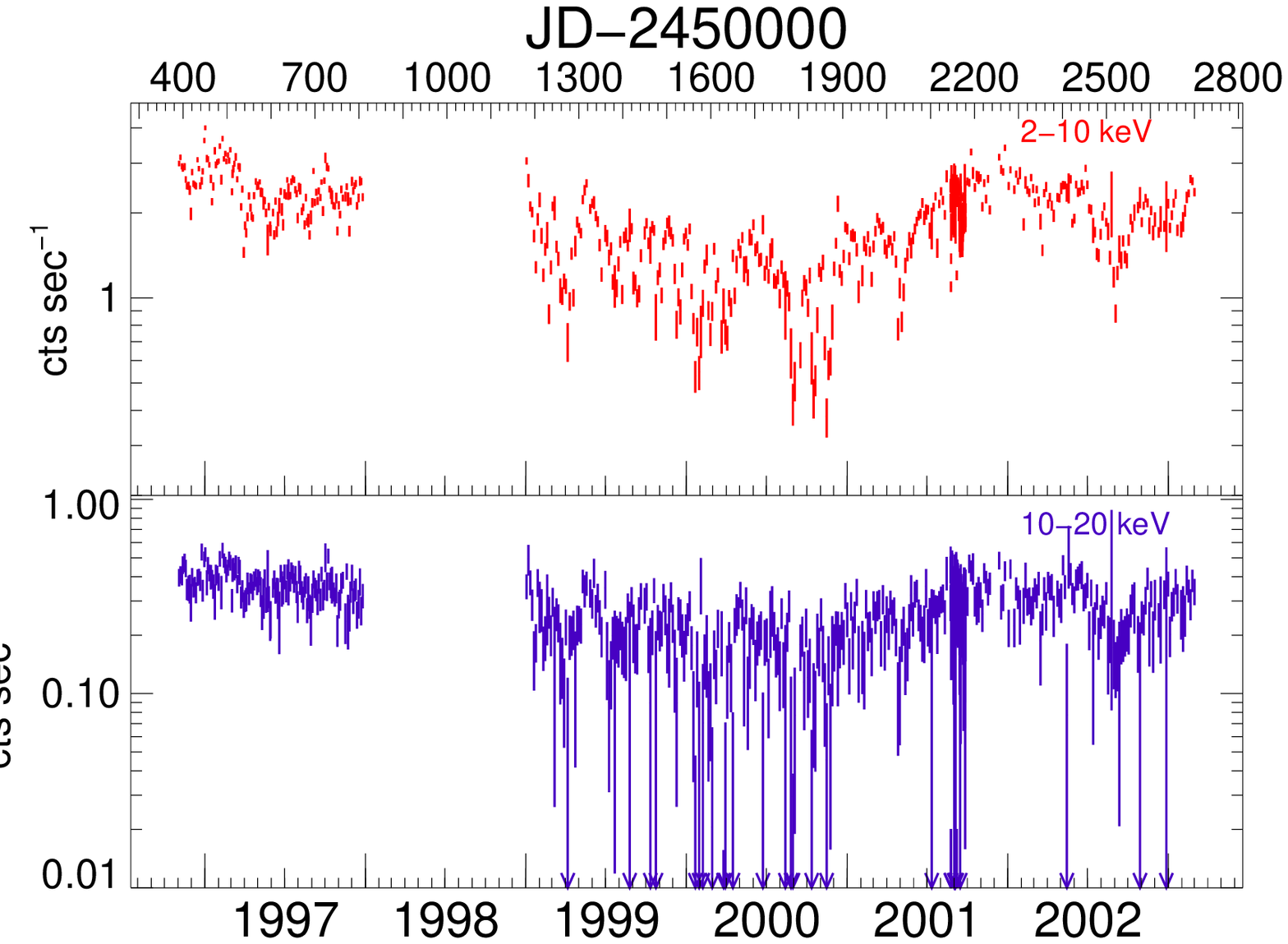}
\includegraphics[width=0.5\textwidth]{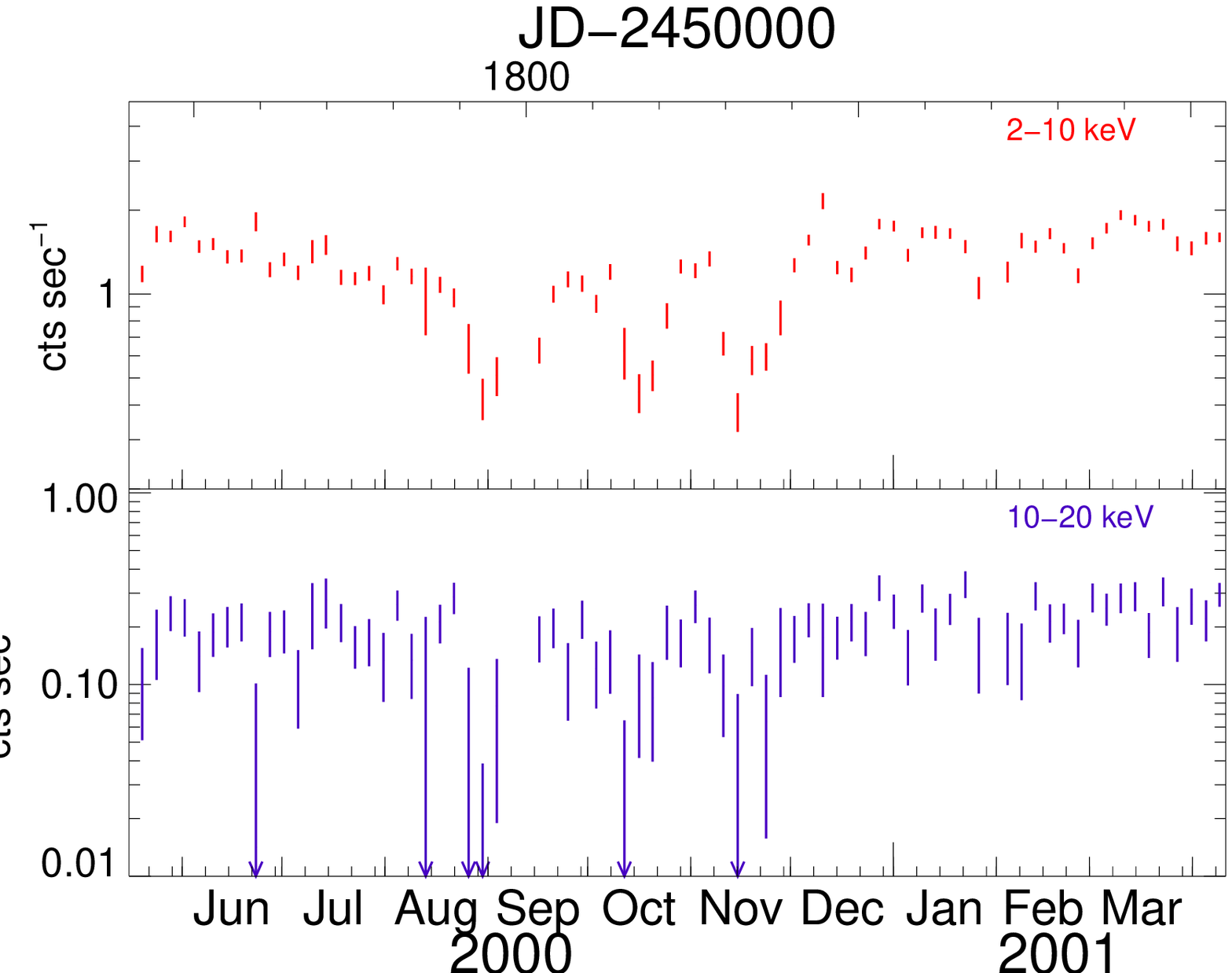}
}
\caption{{\it Left : } Long term \textit{RXTE}-PCA light curves for 2--10\,keV (top panel) and 10--20\,keV (lower panel). {\it Right : }Zoom-in on the \textit{RXTE}-PCA light curves for mid 2000 until 2001 April.}\label{rates}
\end{figure*}

The observations considered in this analysis are all those of Fairall~9 with public \textit{RXTE}/PCA data, a total of 744 pointings spanning the period from late 1996 until early 2003. This gives an average time resolution of about 4\,days with an average pointing length of 1.1\, ks. Only data from the top layer of proportional counter unit 2 were used. The data were reduced as outlined in \citet{wilms:06a}, using {\sc Heasoft 6.10}. For each pointing, we produce a background-subtracted PCA spectrum using the appropriate epoch-dependent background/response files. The count rates quoted below were obtained from the spectra for each pointing. We do not consider any intra-pointing variability in this work.

\section{Results}

\subsection{Light curves and confirmation of dipping}\label{light}

\begin{figure}
\includegraphics[width=\columnwidth]{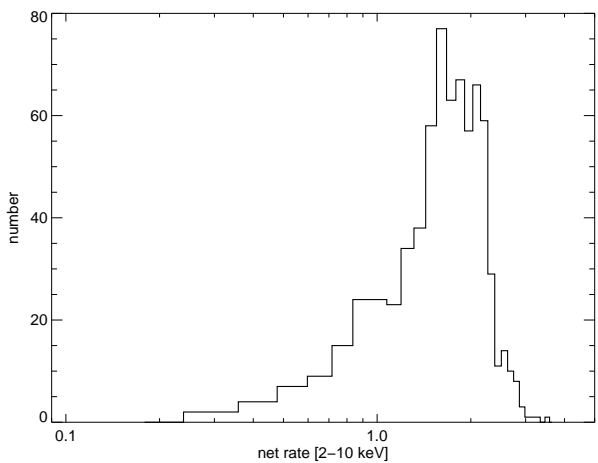}
\caption{Histogram of 2--10\,keV count rate}\label{histogram}
\end{figure}

The long-term light curve in the 2-10\,keV and 10-20\,keV bands is plotted in Figure~\ref{rates} (left).   This light curve displays the unusual behavior already noted in the introduction; long timescale flux modulations punctuated by short and intense dips.   These dips consist of a rapid flux decline by a factor of 2--4 in 5--10 days and then a recovery to the pre-dip level with an entire dip-duration of 10--20 days (Fig.~\ref{rates} right).    The existence of these dips introduces a low-flux tail to the flux distribution, skewing it strongly away from log-normal (Fig.~\ref{histogram}).  To the best of our knowledge, these unusual dips in Fairall~9 have not been commented upon previously.

On longer timescales the source is experiencing a flux decline from early 1999 reaching its low in the fall of 2000 and then recovering to a higher flux value towards the middle of 2001.   There is clearly a greater propensity to dip during the lower-flux state.    The average flux in 2--10\,keV for the object is $(2.16\pm 0.63)\cdot10^{-11}\,\text{ergs}\,\text{s}^{-1}\,\text{cm}^{-2}$.  Both the long term and short term variability seem to be of similar nature in the two different bands.   In particular, while the dips are most clearly seen in the 2--10\,keV light curve due to the higher signal-to-noise, they are also present in the 10--20\,keV light curve. 

\subsection{Spectral Evolution}

To further explore the nature of the variability, we perform flux resolved spectroscopy.   We choose to use the 2--5\,keV flux as our flux discriminator since this will be dominated by the primary power-law rather than a soft excess component or reflection.  We define a ``high state" as $F_{2-5\,{\rm keV}}>1\times 10^{-11}\,\text{ergs}\,\text{s}^{-1}\,\text{cm}^{-2}$ and, by summing the data for all pointing which exceed this flux, form a ``high state spectrum".  Similarly, we define a ``low state spectrum" ($F_{2-5\,{\rm keV}}=(0.5-1.0)\times 10^{-11}\,\text{ergs}\,\text{s}^{-1}\,\text{cm}^{-2}$). A flux of $10^{-11}\,\text{ergs}\,\text{s}^{-1}$ in 2--5\,keV corresponds to a 2--10\,keV count rate of about 1.45 counts per second. A ``dip state spectrum", with a dip defined as a count rate drop of a factor of 1.5 in 2--10\,keV, with respect to the rate right before the dip and the two pointings following the drop, is also defined. The pointings considered to be part of a dip were excluded from high and low states.

Previous studies have shown that, to first order, the spectrum can be well described by a continuum, originating from a comptonizing corona, and cold reflection. We model this scenario using the \texttt{pexmon} model \citep{nandra:07a}, which consists of a power law continuum plus the cold reflection continuum and associated iron K$\alpha$/K$\beta$ fluorescent lines (with a self-consistent strength). The quality of the data does not allow constraints on all parameters in the model, therefore the abundances are assumed to be solar and the inclination of the reflector is fixed to 60\,degrees (the most probable value, assuming a random orientation).  We also include cold Galactic absorption with a column density $N_\mathrm{H}=3.1\times 10^{20}\,{\rm cm}^{-2}$, modeled by \texttt{TBnew}\footnote{http://pulsar.sternwarte.uni-erlangen.de/wilms/research/tbabs/} a newer version of \texttt{TBabs} \citep{wilms:00a} with cross sections set to \texttt{vern} and abundances set to \texttt{wilm}.   

\begin{figure}
\includegraphics[width=\columnwidth]{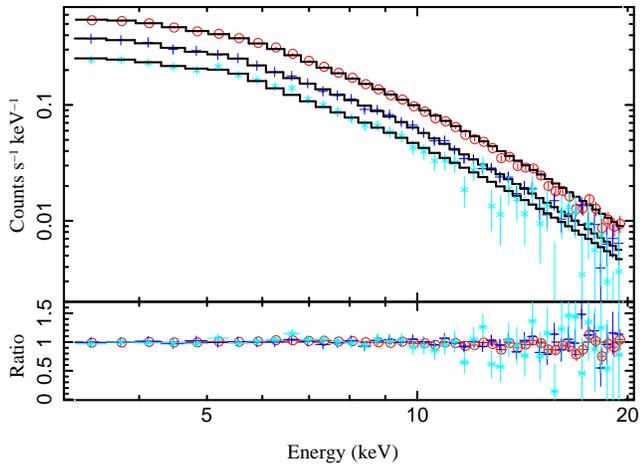}
\caption{Spectra (high state [dots], low state [crosses], dip state [stars]), model [solid line] and residuals for the PCA summed flux resolved spectra for the fit with the cold reflection model.}\label{spectral}
\end{figure}

Table~\ref{spec} shows the spectral fitting results for these three flux-sorted spectra.  The spectra are well described by the model, see Fig.~\ref{spectral}, except for the high state spectrum where a residual line in the iron K region remains. This feature can be modeled with an additional iron line at the redshift of Fairall~9, with an equivalent width of $35\pm8$\,eV.  The uncertainties in the energy of this additional line span both neutral and H-like iron.  Beyond that required by the reflection component, an additional line with this strength is rejected by both the low-state and dip-state spectra.  Thus, it appears that the additional iron line component is exclusively displayed by the high state spectrum.  We comment on the nature of this spectral feature in Section~4.  

\begin{figure}
\vspace{-0.7cm}
\includegraphics[width=1.1\columnwidth]{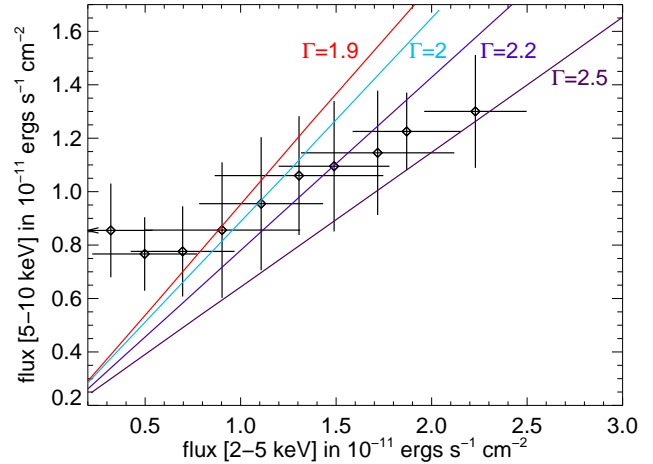}
\caption{Flux-flux plot for the 2--5\,keV and the 5--10\,keV band binned for plotting, showing the non-linearity of the flux-flux relation, as well as model lines for a constant hard component with a flux-variable power law for a set of photon indexes.}\label{col}
\end{figure}

In agreement with previous studies of Seyfert galaxies \citep[e.g., see][]{chiang:00a}, the continuum softens ($\Gamma$ increases) from the low- to the high-state (Fig.~\ref{col}).   We also see a reflection fraction which is inversely proportional to the continuum flux; i.e., the X-ray reflection is consistent with being a constant flux contribution.  The most obvious interpretation is that a significant component of the reflection is from a distant structure. Given that the modulation betwen high- and low-flux states occurs on timescales of $\sim1$\,year, we infer that the distant reflector is situated at least a light year from the central X-ray source.  We note that X-ray reflection from the inner accretion disk is also expected to be present \citep{schmoll:09a,emmanoulopoulos:11a} and, given the error bars on $R$, its presence is consistent with these data.

\begin{table}
\caption{Spectral Parameters for cold reflection fits, see text for details on the model set-up.}\label{spec}
\begin{tabular}{c|c|c|c|c}
Flux State & $\Gamma$ & $R$ & Flux [2--10\,keV] &$\chi^2$/dof \\
\tableline \tableline
high & $2.00_{-0.03}^{+0.03}$ & $1.3_{-0.2}^{+0.2}$ & $2.7\cdot10^{-11}$ & 73.3/42 \\
high w line & $1.90_{-0.03}^{+0.03}$ & $0.5_{-0.2}^{+0.2}$ & $2.7\cdot10^{-11}$ & 16.5/41 \\
low & $1.91_{-0.04}^{+0.04}$ & $1.1_{-0.2}^{+0.3}$ & $1.8\cdot10^{-11}$ & 38.9/42 \\
dip & $1.98_{-0.13}^{+0.14}$ & 2.1$_{-1.0}^{+1.3}$ & $1.2\cdot10^{-11}$ & 32.0/42 \\\hline
\end{tabular}
\end{table}

Although flux resolved spectroscopy is a powerful tool, it does not offer the possibility to study the actual spectral evolution in time. Higher time resolution can be achieved by looking at X-ray flux-flux plots in different bands. Figure~\ref{col} shows the 2--5\,keV/5--10\,keV flux-flux plot; a positive correlation is apparent, although it is clearly non-linear as can be seen by a comparison with a spectral variability model possessing a flux variable power law with a constant hard component (i.e. constant cold reflection).  At high fluxes, the non-linearity appears dominated by a genuine softening of the primary continuum.  This was already seen in earlier studies for Fairall~9 and other Seyfert~1 galaxies \citep{markowitz:01a,sobolewska:09a}.  At the lowest fluxes (corresponding to the dip state) there is a change in the nature of this non-linearity, with the lowest few 2--5\,keV flux points having an almost constant 5--10\,keV flux.   This signals a significant spectral hardening during the deepest parts of the (2--5\,keV) dips. However, the spectrum is never dominated by reflection.

We do note that this non-linearity calls into question the use of $y$-intercepts in extrapolated flux-flux diagrams to determine constant components in AGN spectra \citep[e.g., see][]{noda:11a}.

\section{Discussion and Conclusion}

In this letter, we studied the X-ray variability of Fairall~9 on timescales of a few days to five years.   On the long timescales studied by the flux resolved spectroscopy, a significant fraction of the reflected X-rays appear to maintain a constant normalization, suggesting that potentially a significant fraction of the X-ray reflection might originate from a structure at least a light year across.  The excess iron emission line (above and beyond that associated with the reflected component) in the high-state spectrum suggests that, in this state, there might be an appreciable increase in the amount of Compton-thin material intercepting the primary X-ray continuum.  This material must be out of our line of sight since we see no iron absorption edge in the high-state spectrum.  A possible identification for these feature is iron line emission (fluorescence or radiative recombination) from a radiatively-driven disk wind that forms in the high-flux state. The spectral resolution of the PCA prevents us from constraining the charge state of the additional iron line and hence the ionization state of this additional material.

The most unusual finding is the discovery of strong and sudden dips in the observed X-ray flux from Fairall~9. In the rest of this section, we discuss the nature of these dips. There are two fundamentally different interpretations; transient absorption events (i.e. eclipses) of the inner accretion disk, or a genuine turning-off of X-ray emission from the inner disk.  

We begin by addressing the absorption possibility.  At least upon the immediate onset of a dip, both the 2--10\,keV and 10--20\,keV fluxes drop by similar magnitudes requiring that the absorber be Compton-thick ($N_H>{\rm few}\times 10^{24}\,{\rm cm}^{-2}$).   Suppose that the radius of an absorbing ``cloud" has radius $r=xr_g$ (where $r_g=GM/c^2$ is the gravitational radius) and it is situated $R=Xr_g$ from the black hole.  Assuming Keplerian motion, the velocity of the cloud is $v=X^{-1/2}c$ and hence the eclipse of the inner disk by the cloud with take a duration $\Delta t\sim 2r/v=2xX^{1/2} (r_g/c)$.  For a black hole mass of $M\approx 3\times 10^8\,M_\odot$ \citep{peterson:04a}, this gives $\Delta t\sim 3000xX^{1/2}\,{\rm s}$.  The observed 2--10\,keV dips have duration $\Delta t \approx 2\times 10^6\,{\rm s}$, implying that $xX^{1/2}\approx 700$.  Assuming that the X-ray source has a size of $r_X=10r_g$, we must have $x\gtrsim 10$ ($r\gtrsim 5\times 10^{14}\,{\rm cm}$) in order for the cloud to be able to block the source and create a strong dip, implying $X\lesssim 5000$ and $v\gtrsim 4300\,{\rm km}\,{\rm s}^{-1}$.  Thus, the absorbing clouds are consistent with being at BLR distances.   For a cloud with radius $r\sim 10^{15}\,{\rm cm}$ to be Compton-thick, it must have a density exceeding $n\sim {\rm few}\times10^{9}\,{\rm cm}^{-3}$.  Thus, these absorbing clouds also have a density consistent with that of BLR clouds \citep[e.g., see ][]{korista:00a}

X-ray eclipses by putative BLR clouds have been found recently in some other AGN, notably NGC~1365 \citep{maiolino:10a} and Mrk~766 \citep{risaliti:11a}.  There are some important differences between the Fairall~9 events and the events in these other sources.  Firstly, in NGC~1365, even the non-eclipse spectrum shows significant absorption.  By contrast, the non-dip state of Fairall~9 is notable for being so clean, having {\it no} discernible cold or ionized absorption seen in any of its previous {\it ASCA} \citep{reynolds:97a}, {\it XMM-Newton} \citep{gondoin:01a,emmanoulopoulos:11a} or {\it Suzaku} \citep{schmoll:09a} observations.  Secondly, the eclipsing clouds in NGC~1365 and Mrk~766 are generally Compton-thin with strong evidence that the column density spikes up and then steadily decreases during the event. This led to the notion of BLR ``comets" \citep{maiolino:10a} with high column density heads followed by lower column tails. By contrast, the Fairall~9 clouds must be Compton-thick, at least at the beginning of the dip event. However, a ``comet" scenario would predict a Compton-thick to Compton-thin transition, leading to a recovery of the hard band light curve before the soft band dip ends. The signal to noise of the current data is insufficient to detect such an early hard-band recovery. Future studies with \textit{Suzaku}. \textit{NuSTAR} or \textit{Astro-H} are required to test these predictions.

The presence of well-defined Compton-thick clouds in an otherwise extremely clean environment raises questions about cloud confinement and, more generally, the nature of these clouds \citep{kallman:85a,snedden:07a}. A detailed discussion of the physics of such clouds is beyond the scope of this Letter.   We do however note that the properties of these clouds (Compton-thick and $r\sim10^{15}\,{\rm cm}$) are consistent with the irradiated envelopes of circumnuclear stars \citep[i.e., ``bloated stars",][]{edwards:80a,penston:88a}.   Bloated stars have been ruled out as a viable model for the broad emission lines themselves on the basis of the smoothness of the H$\alpha$ profile \citep{arav:97a,laor:06a}.  However, especially in high-mass systems such as Fairall~9, bloated stars may nevertheless be present and could readily eclipse the X-ray source producing dips.  

Could the dips be intrinsic to the accretion process, i.e., signal a genuine ``turning-off" of X-ray emission from the inner accretion disk?   It will be useful to delineate a few characteristic timescales for the inner regions of this AGN.   The light crossing time of the inner X-ray emitting regions of the accretion disk (say, out to $r=10r_g$) is $t_{\rm lc}\approx 30\,{\rm ks}$, and the dynamical timescale of the disk at $r=10r_g$ is $t_{\rm dyn}\approx 50\,{\rm ks}$.   Assuming an angular momentum transport parameter $\alpha\approx 0.1$, the thermal timescale at this radius is $t_{\rm th}=t_{\rm dyn}/\alpha\approx 500\,{\rm ks}$.   Finally, we consider the viscous timescale, $t_{\rm visc}=t_{\rm th}/(h/r)^2$ where $h$ is the geometric thickness of the accretion disk.   With an average 2--10\,keV flux of $2.2\times 10^{-11}$ergs/sec/cm$^2$ and using a bolometric correction of 50 from \citet{marconi:04a}, we find an Eddington ratio of ${\cal L}\approx 0.15$ for Fairall~9.  For such an Eddington ratio (and assuming a radiative efficiency of $\eta=0.1$), standard accretion disk theory \citep{shakura:73a}, the inner disk is radiation pressure dominated and has a geometric thickness $h=(3{\cal L}/2\eta)r_g\approx 2.3\,r_g$.   Thus, at $r=10\,r_g$, we have $h/r=0.23$ and a viscous timescale of $t_{\rm visc}\approx 9.4\,{\rm Ms}$ (i.e., approximately 110 days).   

The most obvious (albeit dramatic) scenario is one in which the innermost parts of the radiatively-efficient optically-thick accretion disk are destroyed/ejected by a dynamical or thermal instability.  This is hypothesized to occur in radio-loud broad line radio galaxies (BLRGs) as part of the process of jet formation \citep{marscher:02a,chatterjee:09a,king:11a}.  In the BLRGs, a distinct dip in the X-ray flux precedes large radio flares and the creation of a new superluminal knot.   Since, as with Seyferts, the X-ray emission in BLRGs is also dominated by the corona of the inner accretion disk \citep{marshall:09a}, this X-ray/radio connection conclusively demonstrates a link between changes in the inner accretion disk structure and powerful jet ejection events. 
In fact, it is instructive to compare directly Fairall~9 with the BLRG 3C120.  3C120 has a slightly lower-mass black hole, $6\times 10^7\,M_\odot$ for 3C120 compared to $3\times 10^8\,M_\odot$ for Fairall~9 \citep{peterson:04a}.   However, they possess a very similar Eddington ratio, ${\cal L}_{\rm F9}\approx 0.15$ compared to ${\cal L}_{\rm 3C120}\approx 0.11$, derived using the bolometric luminosity for 3C120 from \cite{vasudevan:09a}.  Thus, their accretion disks might be expected to be in rather similar regimes of behavior.   It is possible that the X-ray dips seen in Fairall~9 have the same physical origin as the ejection-related X-ray dips seen in 3C120 but, for some reason, these disk disruption events lead to the creation of powerful jet outflows in 3C120 but not in Fairall~9 \citep[indeed, Fairall~9 has yet to be detected in the radio band, with an upper limit of $<36$\,mJy at 1.4\,GHz, see][]{whittle:92a}.   Either the magnitude/direction of the black hole spin or the magnetic field structure in the inner disk may be the determining factor in deciding whether disk disruptions generate relativistic outflows. For Fairall~9, the black hole spin has already been measured and was found to be moderate \citep{schmoll:09a}. Observations of Galactic black hole X-ray binaries support this idea of viewing the dips in the context of jet ejections, as some were observed to possess very faint jets, undetectable in case of an AGN \citep{gallo:05a,gallo:07a}.



\acknowledgments
\section*{Acknowledgments}
AML and CSR thank support from the NASA Suzaku Guest Observer Program under grant NNX10AR31G. The authors thank the anonymous referee for comments that improved this manuscript.
This research has made use of data obtained from the High Energy Astrophysics Science Archive Research Center (HEASARC) provided by NASA's Goddard Space Flight Center.


\end{document}